\documentclass[journal]{IEEEtran}
\usepackage{blindtext}
\usepackage{graphicx}
\usepackage{hyperref}

\ifCLASSINFOpdf
\else
\fi

\begin{document}
%
\title{Single Chip Self-Tunable N-Input N-Output PID Control System with Integrated Analog Front-end for Miniature Robotics}

\author{\IEEEauthorblockN{Anindya Shankar Bhandari*, Arjun Chaudhuri*, Mrigank Sharad\\}
\IEEEauthorblockA{Department of Electronics and Electrical Communication Engineering\\
Indian Institute of Technology, Kharagpur\\
Email: anindyaorasb@gmail.com, arjunchaudhuri7@ece.iitkgp.ernet.in, mriganksh@gmail.com\\}
\and
\IEEEauthorblockN{*The contribution by these authors are equal}}
\maketitle

\begin{abstract}
In this work, we explore the design of an integrated, low power single chip multi-channel Proportional-Integral-Derivative (PID) controller for emerging miniature robotics, that includes N inputs and N corresponding outputs thereby resulting in N parallel channels in the control system. It includes analog front-end (AFE) and analog PID controllers for PID parameter tuning based on PSO algorithm. The AFE incorporates adaptive biasing to ensure low power. The PSO is optimized with respect to tuning precision, power and area. This makes it attractive for real-time tuning of multiple miniaturized robotic devices with a single PSO tuning algorithm block assigned for the task. For simulation and testing purposes, we take N as $3$ with the channels being defined by their application-ends or plants, namely: dc motor, temperature sensor and gyroscope.


\end{abstract}


%
\IEEEpeerreviewmaketitle

\section{Introduction}
Miniaturisation of integrated circuits has been the technological motivation behind many of the recent advances in the field of layout, board and circuit optimisation at different levels of abstraction. Beginning from the algorithmic approach at the very top, the journey through the architectural, gate and transistor level design approaches in that hierarchy has brought about notable revolutions in the chip size and power. Technology (channel length, $V_{DD}$, et. al.) and interconnect scaling have been the crust of many research routines to enforce the prediction of the Moore's law. Control systems and engineering is one of the very concise applications where the scaling of chip size and power has been integrated with the sustainability of the stability of a plant, by essence of an analog proportional-integral-derivative (PID) controller. Self-tunable PID controller has been in use in recent control systems because of its autonomous nature of correcting the variation in the stability parameters of a plant or system, namely the $K_p$, $K_i$ and $K_d$ values. Particle Swarm Optimisation (PSO) algorithm is a popular algorithm that has been implemented in tuning the PID controller in this work. This tuning algorithm has been efficiently put to use with the maximum tuning precision and minimum number of converging iterations. The chip has been designed with the analog front end (AFE) design of the PID controller and the Pulse Width Modulation (PWM) driver, which will be driving the motor of the miniature robot, both on a single chip, making the design of the IC highly integrated and occupying minimum area. The overview of the design proposal is given in Figure $1$.\newline
In the next section, the previous research work on self-tunable PID controller and its components will be detailed along with the literature on the PWM driver design and custom DSP. Then, we will be exploring the methodologies of designing the PSO tuning section, the AFE and the PWM driver, highlighting the compatible connections among the different functioning blocks which will be integrated on the same chip or die area. Subsequently, the possible future endeavours, which this research work can open the doors to, will be discussed briefly. 

\begin{figure}[bhp]
\includegraphics[width=\columnwidth]{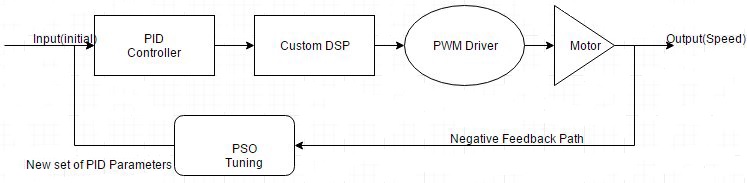}
\caption{Holistic Block-level View of a Single Channel}
\end{figure}

\begin{figure}[bhp]
\includegraphics[width=\columnwidth]{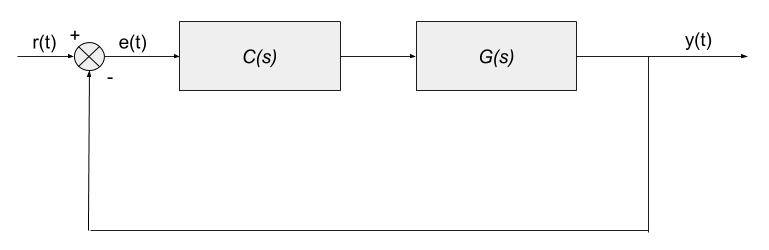}
\caption{PID Block Diagram}
\end{figure}

\begin{figure}[bhp]
\includegraphics[width=\columnwidth]{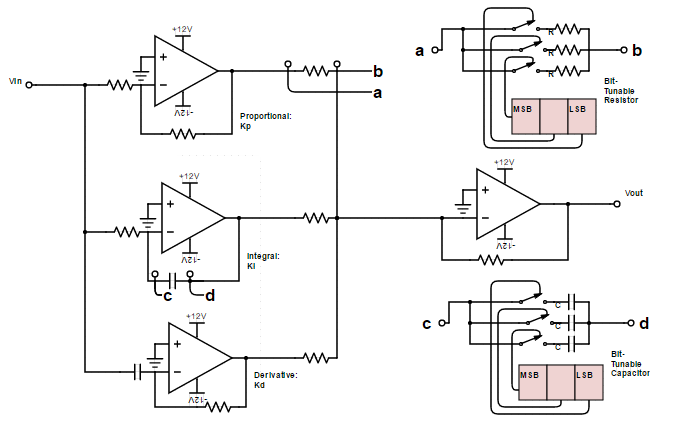}
\caption{Circuit Schematic of the PID controller with Bit-tunable resistors and capacitors}
\end{figure}

\begin{figure}[bhp]
\includegraphics[width=\columnwidth]{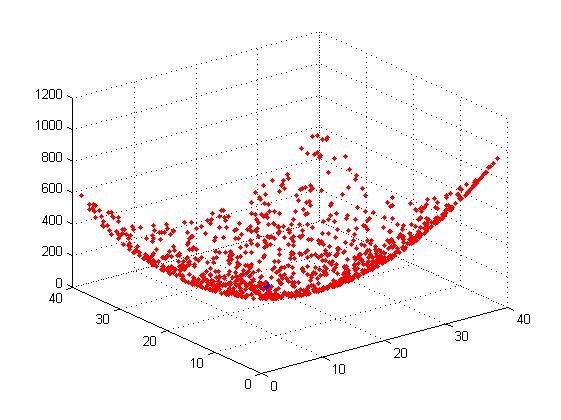}
\caption{Convergence of the Swarm of Nodes or Particles}
\end{figure}

\begin{figure}[bhp]
\includegraphics[width=\columnwidth]{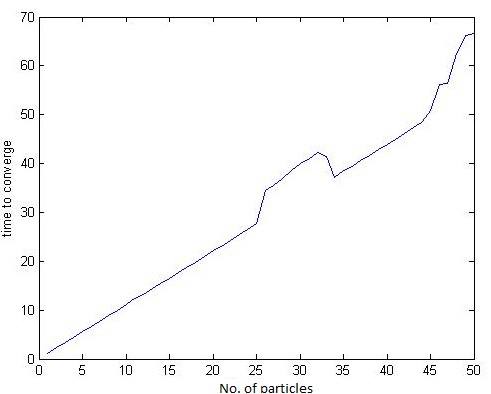}
\caption{Convergence time vs No. of Particles}
\end{figure}

\begin{figure}[bhp]
\includegraphics[width=\columnwidth]{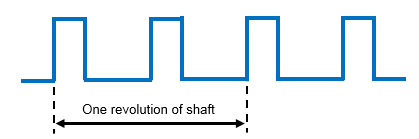}
\caption{Tachometer Waveform Showing 2 Pulses per Revolution}
\end{figure}

\section{Literature Review}
In this research, emphasis had been put on the previous work that has already been done in the field of self-tunable PID controller, so that advancement in terms of novelty can be made in area and power of the chip which will also accommodate the PWM driver of the motor.\newline
In $[1]$, self tuning Analog PID controller design has been discussed along with the benefits of analog self tuning PID controller over the digital PID controller and provides the comparison with hand tuned solutions. However, in this work, the technique has been improved to give maximum tuning and bit precision.\newline
The digital version of the PWM driver (DPWM: digital pulse width modulation) for the PID controller using digital DC-DC converters has been proposed in $[2]$, which has led to the motivation of designing the analog version of the PWM driver, therefore eliminating digital blocks and reducing power dissipation.\newline
$[3]$ gives a detailed description on the self-tuning PID controllers and is a nice reference for the theory of the same.\newline
In $[4]$ and $[5]$, the PID controller tuning methods and design specifications and parameter selection reference for PSO tuning algorithm, respectively, have been discussed. These have led to a better understanding of the fundamental working of the self-tuning PID controller using the PSO algorithm and further optimising the approach for less number of iterations to converge to the desired set of values for $K_p$, $K_i$ and $K_d$.\newline
Zieglar-Nichols method for optimally finding the $K_p$, $K_i$ and $K_d$ values of a PID controller was proposed in $[6]$.\newline 
The reference to the PSO algorithm was taken from $[7]$, where the algorithm was proposed for the first time.\newline 
In the works of $[13]$,$[14]$,$[15]$,$[16]$ and $[17]$, we find similar applications of FPGA or non-FPGA based control systems for micro robotics in low power ICs. All of these chips have focussed on a single input channel or plant which can be improved upon considerably following the idea we have proposed in this paper.\newline 
In $[18]$, nuero-inspired PID controllers have been utilised for robotics applications. However, as explained in our work, the sharing of PID blocks along with a back-up PSO algorithm for finer and more reliable tuning presents a step further for similar research.\newline 
$[19]$ and $[20]$ present low-power and low-area control systems on chip for robotic control plants for large scale sensor networks (FPGA or non-FPGA-based) but neither of these implements the tuning approach detailed in this paper that not only minimises the number of PID controllers for the closed loop control but also introduces a 2-tier mechanism of control transfer.
\section{Particle Swarm Optimization for PID Tuning}
As stated before, we used the Particle Swarm Optimization technique to tune the parameters $K_p$ and $K_d$ of the PID Controller. In this section, we shall get into the details of our approach.\\
Particle swarm optimization (PSO) is a population-based stochastic optimization method. It was developed by Dr. Russell Eberhart and Dr. James Kennedy in 1995. It was inspired by the behaviour of birds flocking as a group, from a psychological perspective. It shares many similarities with evolutionary computation techniques such as Genetic
Algorithms $[8]$.\\
It is essentially a computational method that optimizes a problem by iteratively trying to improve a candidate solution (which starts off as an initial guess - However this may be modified according to the application) with regard to a given measure of quality (comparing values against a cost function). It solves a problem by having a population of candidate solutions (a random initialization of particles in space), here dubbed particles, and moving these particles around in the search-space according to simple mathematical formulae
over the particle's position and velocity. Each particle's movement is influenced by its local best known position, but is also guided toward the best known positions in the search-space, which are updated as better positions are found by other particles. This is expected to move the swarm toward the best solutions, just like a swarm of birds, where the birds follow the bird closest to the target (food) $[9]$.\\
A proportional integral derivative controller is a control loop
feedback mechanism (controller) often used in industrial control
systems. A PID controller continuously calculates an error value as the difference between a desired value and a measured process
variable.\\
Tuning of PID controllers is essential for the proper functioning of the controller. Among conventional tuning algorithms, the most well known is the Ziegler Nichols method. However, it does not always return a desirable result, and often results in an overshoot$[10]$.\\
Both genetic algorithms and particle swarm optimization have been utilized for this purpose, however we chose to use PSO. The PID block diagram, depicting the feedback path, is shown in Figure $2$.\\

\begin{figure}[bhp]
\includegraphics[width=\columnwidth]{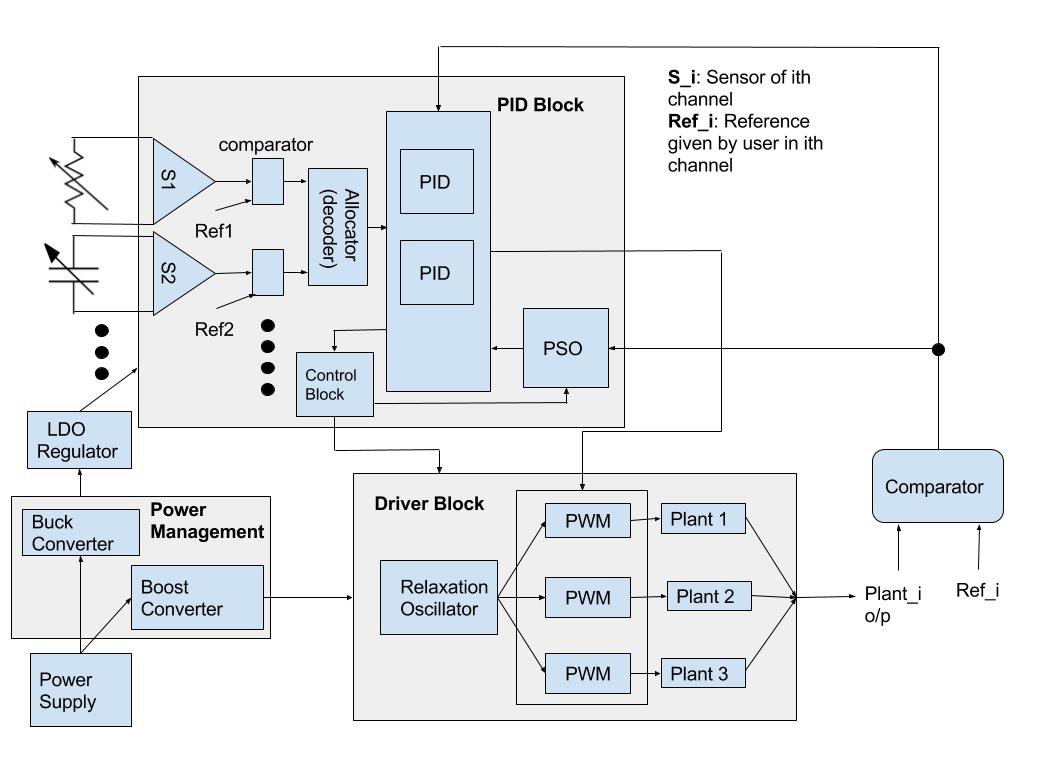}
\caption{Block View of 2-level Shared Multiple Input Multiple Output PID Control System}
\end{figure}

The PID Controller Transfer Function is given as follows:
$$C(s)=K_p+\frac{K_i}{s}+K_ds$$
\subsection{Implementation}
In our implementation of PSO for the PID controller, we used a 3
dimensional position-velocity model, of which individual dimensions of
the final position represented the $K_p$, $K_i$ and $K_d$ values. We
initialized and worked with 50 particles in space. The model was run
for 50 iterations.
\subsection{Initializing the model}
The model was initialized with the number of particles and iterations
both as 50, the current position and velocities randomly initialized,
and the current fitness calculated using the objective function
described in the next section.
\subsection{Objective Function}
The objective or `fitness' function used for the PSO was
$F=\beta*Error+\alpha*sys_{overshoot}$ where $\beta$ and $\alpha$
represent the weightage given to the error and to the overshoot. In
our implementation, we weighed these equally.
\subsection{Updating the model}
The model was updated by first calculating the current velocity based
on the following equation:
 $v_{new}= w\times v_{current} + c_1(R_1 .*
(x_{local\ best}-x_{current}))+ c_2 (R_2.* (x_{global\
best}-x_{current}))$\\
Here $R_1$ and $R_2$ are the 2 randomly
initialized matrices mentioned before, and the $.*$ operation
represents element wise multiplication. Using this updated velocity,
the current position is updated as\\
 $$x_{current}=x_{current}+v_{new}$$.
The PSO algorithm iteratively converges to a solution. This has been diagrammatically depicted in Figure 4 [16]. Figure 5 shows how the convergence time varies with the selection of particles.

\ifCLASSOPTIONcaptionsoff
  \newpage
\fi

\section{Analog Front-end Design of the PID Controller}
As shown in Figure $3$, the PID controller has been designed to satisfy minimum area requirements, consisting of three major blocks of operation:
\begin{itemize}
    \item Proportional block: Output of this block is $V_{out1}=-\frac{R_2}{R_1}.V_{err}(t)$, where $K_p=-\frac{R_2}{R_1}$.
    \item Integral block: Output of this block is $V_{out2}=-\frac{1}{R_i.C_i}.\int_{}^{} V_{err}(t)dt$, where $K_i=-\frac{1}{R_i.C_i}$.
    \item Derivative block: Output of this block is $V_{out3}=-R_d.C_d.\frac{dV_{err}(t)}{dt}$, where $K_d=-R_d.C_d$.
\end{itemize}
Figure $3$ shows the circuit level schematic of the analog PID controller.
A fourth operational amplifier with $R_1$ and $R_2$ is placed just before the node where the output is taken to amplify the signal coming into it. 

\begin{table*}[t]
\centering
 \begin{tabular}{||c | c | c | c | c | c ||} 
 \hline
 Device & Input Swing & Output Swing & Mapping & Notes & URL\\ [0.5ex] 
 \hline
 Temperature Sensor & 0 - 5 V & 0.25 - 4.75 V & 22.5 mV/C & 200C Temp Span & \href{https://cdn.hackaday.io/files/5059216444256/AD22100.pdf}{Here}\\
 \hline
 Gyroscope & 0 - 5 V & 0 - 5 V & 107.42 mV/(rad/s) & -23.27 to 23.27 rad/s & \href{http://www.pieter-jan.com/node/7}{Here}\\
 \hline
 Tachometer & 0 - 5 V & 0 - 5 V & $f_{out} = \frac{RPM \times N}{60}$ & N is shaft teeth & \href{http://www.ni.com/white-paper/3230/en/}{Here}\\
 \hline \hline 
\end{tabular}
\\
\textbf{Table 1: Parameters for the various physical systems connected to each channel of the setup}
\end{table*}

\subsection{Implementation}
All the four op-amps have been designed to be inverting in nature. The resistors and capacitors are bit-tunable to the precision of 4 bits so that they can be programmed directly by the output (${K_p,K_i,K_d}$) values coming from our PSO algorithm which are digital in nature. $K_p$, $K_i$ and $K_d$ are stored in 4-bit registers each and each of these 12 bits programs the bit-tunable resistors and capacitors to yield the desired PID control parameters in the analog circuit designed previously and achieve the desired set-point (here, desired speed of the motor of the miniaturised robot which our model plant simulates) eventually.

\subsection{Design of the Bit-tunable Components}
Bit-tunable resistors and capacitors have been designed by simple switches connected to the corresponding bits of the register storing the binary representations of the desired resistor and capacitor values, as seen in Figure $3$.

\section{Detection of Plant parameter disturbance and multiplexing}
In our experiment, we have sixteen channels, three of which have been shown for better clarity-tachometer channel, gyroscope channel and temperature-sensor channel. In general, many more channels can be implemented.  At the outputs of each of the sixteen channels' plants, an analog comparator is present that trivially compares the plant output (or its voltage equivalent) to the reference level given by the user and generates a high or low signal depending on the result of the comparison. This high/low signal triggers an analog multiplexer or allocator (comprising the decoder) which assigns a certain PID controller block to that channel where the disturbance has occurred. Here, the 2-tier threshold tuning begins:

\begin{itemize}
    \item Case 1: If the PID controller is able to reduce the error in the signal level within a stipulated time $t_o$, then post-tuning, the system reverts to normal working condition and the PID controller becomes idle.
    \item Case 2: If the PID controller is unable to reduce the error in the signal level within the stipulated time $t_o$, it invokes the PSO block to re-configure its $k_p,k_i,k_d$ values and tune it. This in effect reduces the error in the signal with respect to the reference level.
\end{itemize}
This switching of the control from the PID to the PSO and back to the PID is monitored by the Control Box.

\begin{figure}[bhp]
\includegraphics[width=\columnwidth]{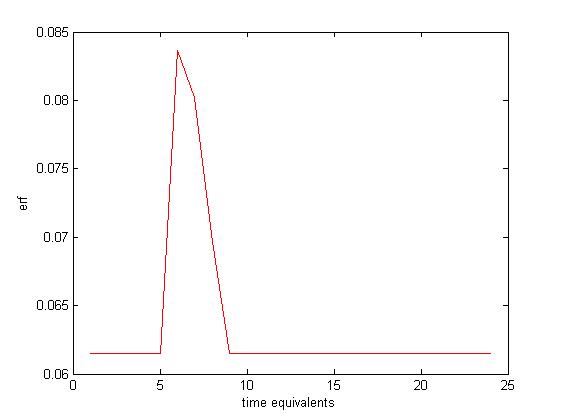}
\caption{Momentary Pulse, Removed at t=8}
\end{figure}
\begin{figure}[bhp]
\includegraphics[width=\columnwidth]{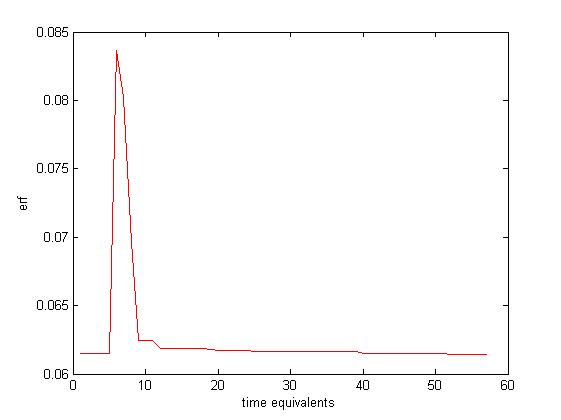}
\caption{Sustained Disturbance, tuned by PSO}
\end{figure}
\begin{figure}[bhp]
\includegraphics[width=\columnwidth]{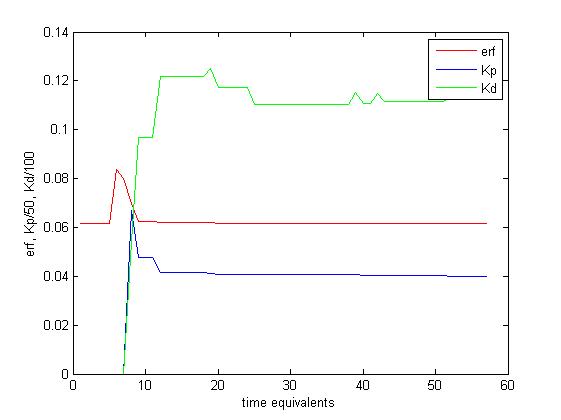}
\caption{Sustained Disturbance, Showing Kp, Kd and erf}
\end{figure}

\section{Physical Devices}
For our experiment, in each channel we used one of 3 (here, $N=3$ for experimental convenience) physical devices (or "plants") (devices that require the transformation of some form of physical energy to electrical energy). They are listed as follows:\newline
(1) Temperature Sensor: A temperature sensor ($[11]$) works by converting detected external temperatures to a range of voltages, each level of voltage denoting a particular value of external temperature.\newline
(2) Gyro-Sensor: A Gyroscope measures the angular velocity of a device, converting those values to voltage levels.\newline
(3) Tachometer: A tachometer measures the speed of a system or device, representing them as voltage levels, based on revolutions of the shaft (Figure $6$, taken from $[12]$). \\
For the specifications-range of input/output value(s) and the conversion or offset factor(s), refer to Table $[1]$.

\section{Experimental Setup}
We set up our system with multi-channel inputs and outputs as can be observed in Figure $7$. \\

Our system is fed multi-channel inputs from different kinds of physical systems, such as temperature sensors, gyroscopes and tachometers.\\
These physical systems need to have compatible input-output voltage swings, and we also require to know the conversion factor to map the physical dimensions of the instrument (such as temperature or speed) into Voltage, for practical tuning by the PID controller. A table of these values can be seen in Table $1$.\\
If the PID controller detects an anomalous input at any of the input channels, the PSO-based PID tuner shall attempt to optimize the system by minimizing the error introduced by the fluctuation or change in the input values, by a control loop feedback system. The PSO-based method allows for a ``self-tuning" PID controller, which is an integral part of miniature robotics based on multi-channel information pathways.\\
In [1], Varun Aggarwal et al attempted to do the same thing, but based on a single channel (a plant model), and realized its implementation on a Field Programmable Analog Array (FPAA). Our system builds on this by utilizing individual analog components and allowing for multi-channel inputs, thereby a marked evolution of the Analog Self Tuning PID domain, and a big step towards the further miniaturization of robotics-related components.

\section{Results}
In this paper, the compact design of an analog multi-channel PID controller with integrated analog front end and PWM driver, all on the same chip, was proposed. The PSO was tuned to a bit-precision of 4 bits and it was shown that amidst the trade-off between time required for a fixed number of nodes or particles (Figure $5$) and the corresponding bit number required to represent the PID controller parameters, an optimum point was obtained. This point gave us the necessary bit-precision for designing the storage registers and the bit-tunable resistors and capacitors. No ADCs were used in the entire system set-up that brought down the cost of power consumption by an appreciable amount. The novelty of the idea behind this work lies in the fact that the entire design is mostly analog with implementation of basic circuits in it, yet minimising the area and power to a large extent. This was possible owing to the utilisation of bit-variability of the PSO design and the use of a mathematical function to map the motor speed to a set of voltages used for reference by the PSO and the PID controller.\\
We also obtained the following graphs. Here `t' represents a time equivalent, where 1 time equivalent is equal to the time taken for one iteration of the PSO tuning algorithm. Figure $8$ shows how the PID Controller attempts to stabilize a pulse switched on at t=6 and switched off at t=8. If the error is stabilized within a specified duration (t=3 time equivalents, in our case) period, the PSO tuner does not activate.\\
In case the PID fails to stabilize the input within that period, the PSO tuner activates. Figure $9$ shows how the error converges using the tuning algorithm.\\
Finally, Figure $10$ shows the variation of Kp and Kd during the tuning process.\\

\section{Conclusion and discussion}
By means of multiplexing, a single PSO tuner was able to tune the three channels, thus economically utilising power for multi-channel PID control. The multiple input, multiple output system thus employed a single PSO tuning algorithm block to tune all the three PID controllers, one at a time. As the PSO tuner converges in a maximum of 50 iterations, the waiting time for the remaining two channels will be extremely small while one channel is being tuned. This incorporates tractability into the system and proposes a convenient way of handling all the N channels in a pseudo-parallel manner.  

\section{future work}
Miniature robotics and similar applications will find this multi-channel, multi-user optimisation technique, involving reduction in chip area by adding major functionalities in the same chip or die area, progressive in further reduction in their sizes and masses. Research is ongoing on improving the PSO performance by reducing the power consumed by the PSO block alone by bringing down the number of iterations further for the same number of nodes or particles. Faster convergence (convergence of the swarm portrayed in Figure $4$) will result in less switching of the register values resulting in lower power consumption by the digital circuit constituted within the PSO tuning block. In the analog front end design, reduction in hardware size for the bit-tunable components is quintessential for future advancement of this particular area. Additionally, scaling down of $V_{DD}$, and even in the sub-threshold regime, is a major factor that will decrease the amount of power dissipated. Improvement on the already proposed method will enhance user-handling capacity of the N-input N-output system and will be able to accommodate N plants in it, where $N>3$. 

\section{References}
1. V. Aggarwal, Meng Mao, and U.-M. O’Reilly. A self-tuning analog proportional-integral-derivative (pid) controller. In Adaptive Hardware and Systems, 2006. AHS 2006. First NASA/ESA Conference on, pages 12 – 19, 2006.\newline
2. Sun A., Tan M., Siek L., Segmented Hybrid DPWM and Tunable PID Controller for Digital DC-DC Converters, In: Proceeding of the IEEE International Symposium on Next-Generation Electronics (ISNE’10), 2010, 154-57.\newline
3. Vladimír Bobál, Josef Böhm, Jaromír Fessl and Jiří Macháček. Digital Self-tuning Controllers, Springer Books. http://www.springer.com/in/book/9781852339807 \newline
4. P. Cominos, N. Munro, PID controllers: recent tuning methods and design to specifications, IEE Proceedings––Control Theory and Applications 149 (1) (2002) 46–53.\newline
5. Y. Shi and R. Eberhart. (1998). Parameter Selection in Particle Swarm Optimization. Proceedings of the Seventh Annual Conf. on Evolutionary Programming, March 1998.\newline
6. Ziegler JG, Nichols NB. Optimum Settings for Automatic Controllers. ASME. J. Dyn. Sys., Meas., Control. 1993;115(2B):220-222. doi:10.1115/1.2899060.\newline
7. Kennedy, J., and Eberhart, R. C. (1 995). :Particle Swarm Optimization; Proc. IEEE International Conference on Neural Networks (Perth, Australia), IEEE Service Center, Piscataway, NJ, IV: 1942-1948. \newline
8. Swarm Intelligence Website. http://swarmintelligence.org\newline
9. Wikipedia.org\newline
10. M. Solihin, L. Tack, and M. Kean, “Tuning of PID controller using particle swarm optimization (PSO),” Proceeding of the International conference on Advanced Science, Engineering and Information Technology, pp. 458-461, January 2011\newline
11. Analog Devices Manual for AD22100, A Voltage Output Temperature Sensor with Signal Conditioning.\newline 
12. National Instruments, Decoding Tachometer Signals Using CompactRIO and LabVIEW FPGA. http://www.ni.com/white-paper/3230/en/ \newline
13. A. Ferreira, J. Agnus, N. Chaillet and J. M. Breguet, "A smart microrobot on chip: design, identification, and control," in IEEE/ASME Transactions on Mechatronics, vol. 9, no. 3, pp. 508-519, Sept. 2004.
doi: 10.1109/TMECH.2004.834646 \newline
14. L. Arcese, M. Fruchard and A. Ferreira, "Nonlinear modeling and robust controller-observer for a magnetic microrobot in a fluidic environment using MRI gradients," 2009 IEEE/RSJ International Conference on Intelligent Robots and Systems, St. Louis, MO, 2009, pp. 534-539.
doi: 10.1109/IROS.2009.5354600 \newline
15. Y. A. Chapuis, L. Zhou, Y. Fukuta, Y. Mita and H. Fujita, "FPGA-Based Decentralized Control of Arrayed MEMS for Microrobotic Application," in IEEE Transactions on Industrial Electronics, vol. 54, no. 4, pp. 1926-1936, Aug. 2007.
doi: 10.1109/TIE.2007.898297 \newline
16. M. Karpelson, R. J. Wood and G. Y. Wei, "Low power control IC for efficient high-voltage piezoelectric driving in a flying robotic insect," VLSI Circuits (VLSIC), 2011 Symposium on, Honolulu, HI, 2011, pp. 178-179.\newline
17. R. Casanova, A. Dieguez, A. Sanuy, A. Arbat, O. Alonso, J. Canals, and J. Samitier,
“An ultra low power IC for an autonomous mm3-sized microrobot,” in Proc. IEEE
Asian Solid-State Circuits Conference, ASSCC ’07, pp. 55–58, 2007.\newline
18. Jimenez-Fernandez A, Jimenez-Moreno G, Linares-Barranco A,
Dominguez-Morales MJ, Paz-Vicente R, Civit-Balcells A. A NeuroInspired
Spike-Based PID Motor Controller for Multi-Motor Robots
with Low Cost FPGAs. Sensors. 2012; 12(4):3831-3856. 
\newline
19. Robomote: a tiny mobile robot platform for large-scale ad-hoc sensor networks-By G.T. Sibley, M.H. Rahimi and G.S. Sukhatme. resenv.media.mit.edu/classarchive/MAS965/readings/sibley2002.pdf \newline
20. F. J. Lin, D. H. Wang, and P. K. Huang, “FPGA-based fuzzy sliding mode
control for a linear induction motor drive,” Proc. Inst. Elect. Eng.—Elect.
Power Appl., vol. 152, no. 5, pp. 1137–1148, Sep. 2005. \newline

\end{document}